\documentclass[pdflatex]{article}
\usepackage{waspaa19,amsmath}
\usepackage{graphicx}
\usepackage{bm}
\usepackage{url}
\usepackage{amssymb}
\usepackage{cite}

\newcommand{\dataseturl}{\url{https://github.com/YumaKoizumi/ToyADMOS-dataset}}

\makeatletter
\def\Hline{
  \noalign{\ifnum0=`}\fi\hrule \@height 2.\arrayrulewidth \futurelet
  \reserved@a\@xhline}
\makeatother

\title{T\lowercase{oy}ADMOS: 
A Dataset of Miniature-Machine Operating Sounds for Anomalous Sound Detection
}

\name{Yuma Koizumi${}^{1}$, Shoichiro Saito${}^{1}$, Hisashi Uematsu${}^{1}$, Noboru Harada${}^{1}$ and Keisuke Imoto${}^{2}$}
\address{
$1$: NTT Media Intelligence Laboratories, Tokyo, Japan\\
$2$: Ritsumeikan University, Shiga, Japan\\
}

\begin{document}
\ninept
\maketitle

\begin{abstract}
This paper introduces a new dataset called ``ToyADMOS'' designed for anomaly detection in machine operating sounds (ADMOS). To the best our knowledge, no large-scale datasets are available for ADMOS, although large-scale datasets have contributed to recent advancements in acoustic signal processing. This is because anomalous sound data are difficult to collect. To build a large-scale dataset for ADMOS, we collected anomalous operating sounds of miniature machines (toys) by deliberately damaging them. The released dataset consists of three sub-datasets for machine-condition inspection, fault diagnosis of machines with geometrically fixed tasks, and fault diagnosis of machines with moving tasks. Each sub-dataset includes over 180 hours of normal machine-operating sounds and over 4,000 samples of anomalous sounds collected with four microphones at a 48-kHz sampling rate. 
The dataset is freely available for download at \dataseturl.
\end{abstract}




\section{Introduction}
\label{sec:intro}

Since anomalies might indicate faults or malicious activities, prompt detection of anomalies may prevent such problems. Microphones have been used as sensors to detect anomalies, referred to as anomaly detection in sounds (ADS) \cite{Koizumi_2018_IEEE_ADS} or acoustic condition monitoring \cite{Heinicke2015}, in many applications such as audio surveillance \cite{Clavel_2005,Valenzise2007,Ntalampiras_2011,Foggia_2016}, machine-condition inspection, and fault diagnosis \cite{Yamashita_2006,Koizumi_2017_ADS,sniper}. A recent advancement in this area is the use of deep learning \cite{Marchi_2015,Kawaguchi_2017_MLSP,Kawachi_2018,Koizumi_2019_BU,Kawachi_2019,adaflow}: an autoencoder (AE) \cite{Marchi_2015,Kawaguchi_2017_MLSP,Koizumi_2019_BU}, variational AE \cite{Kawachi_2018,Kawachi_2019}, and/or flow-based model \cite{adaflow} are used to calculate the anomaly score.

A large-scale dataset is essential for successfully training and fairly evaluating a deep neural network (DNN)-based system. Therefore, the existence of freely available large-scale datasets often accelerates related research in this domain. For example, the accuracy of computer-vision tasks has rapidly increased thanks to large-scale datasets such as ImageNet \cite{ImageNet}. Large-scale datasets have also contributed to recent advancements in speech and acoustic signal processing such as the Wall Street Journal (WSJ0) speech corpus \cite{WSJ0} for automatic speech recognition and the VCTK corpus \cite{VCTK} for text-to-speech synthesis. In audio-event detection and scene-classification tasks, two large-scale datasets, AudioSet \cite{AudioSet} and the Freesound dataset \cite{FreeSound}, have been published and used in several tasks of the Detection and Classification of Acoustic Scenes and Events (DCASE) challenge \cite{DCASE2017,DCASE2016}.

Unfortunately, to the best of our knowledge, no large-scale datasets are freely available for ADS. One reason is anomalous sounds occur far more rarely than normal sounds, resulting in anomalous sounds being difficult to collect. Thus, surveillance tasks, such as gunshot detection, are trained and evaluated on small-scale datasets \cite{rare_01,rare_02}. Machine-condition inspection and fault-diagnosis tasks hardly have even small datasets. Thus, an ADS system has been evaluated by using a synthetic anomalous sound dataset \cite{Koizumi_2018_IEEE_ADS} instead of collecting anomalous sounds by deliberately damaging expensive machinery. To fairly evaluate systems for anomaly detection in machine operating sound (ADMOS), we believe that a freely available dataset is necessary.

\begin{figure*}[ttthhhhhh] 
\centering 
\includegraphics[width=150mm,clip]
{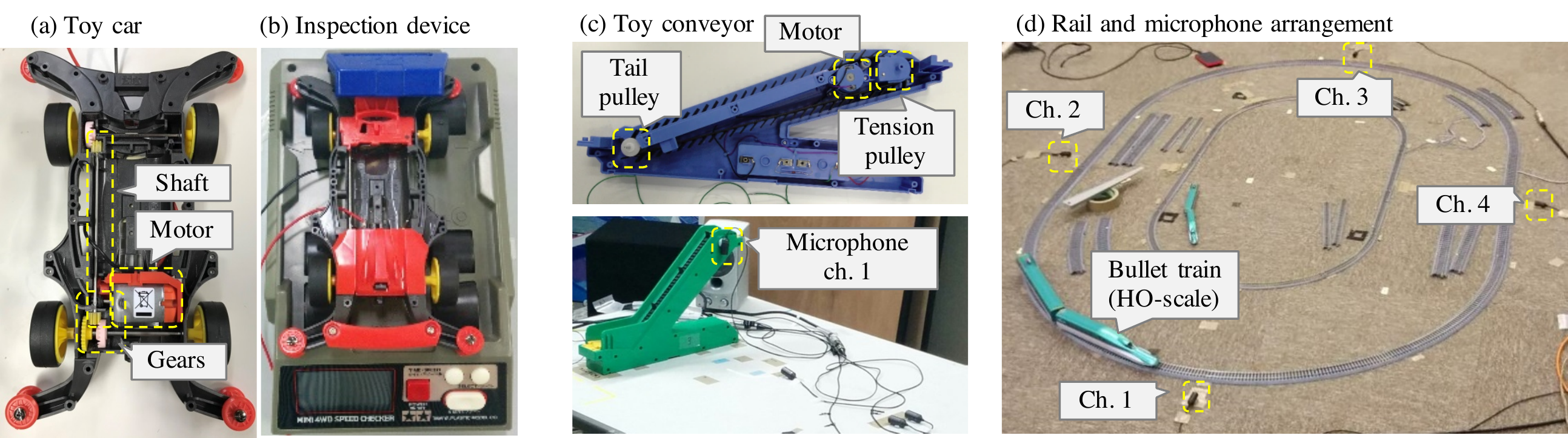} 
\caption{Images of toys, parts, and microphone arrangement} 
\label{fig:mini4wd}
\end{figure*}

This paper introduces a new dataset called ``ToyADMOS'' designed for training and testing ADMOS systems. We collected normal and anomalous operating sounds of miniature machines by deliberately damaging their components. Since miniature machines can be installed in an acoustic laboratory, recording conditions can be controlled. With this advantage, we designed the ToyADMOS dataset to be used for not only basic unsupervised-ADMOS \cite{Koizumi_2018_IEEE_ADS} but also multiple advanced tasks such as domain adaptation \cite{adaflow}, noise reduction \cite{kawaguchi2019}, data augmentation, and few-shot learning of anomalous sounds \cite{sniper}. The ToyADMOS dataset has the following characteristics:
\begin{itemize}
\item It is designed for three ADMOS tasks: product inspection (toy car), fault diagnosis for a fixed machine (toy conveyor), and fault diagnosis for a moving machine (toy train).
\item Machine-operating sounds and environmental noise are individually recorded for simulating various noise levels.
\item All sounds are recorded with four microphones for testing noise reduction \cite{kawaguchi2019} and/or data-augmentation techniques such as mix-up.
\item In each task, multiple machines of the same class are used; each machine belongs to the same class of toys but has a different detailed structure ({\it cf.} Sec \ref{sec:detail_sub_dataset} and supplemental document available on the dataset web-page). Since the collected operating sounds have variations depending on individual differences,
the dataset can be used for testing domain-adaptation techniques to absorb individual differences and/or changes in noise level \cite{adaflow}.
\item Each anomalous sound was recorded several times for testing few-shot learning-based ADMOS for obtaining the characteristics of anomalous sounds from only a few samples \cite{sniper}.
\item The released dataset consists of over 180 hours of normal machine-operating sounds and over 4,000 samples of anomalous sounds collected with four microphones at a 48-kHz sampling rate for each task.
\end{itemize}
This dataset is freely available for download at \url{https://github.com/YumaKoizumi/ToyADMOS-dataset} and the license and some use of the dataset are also available on the webpage.

\section{Dataset overview}
\label{sec:design}
\begin{figure*}[ttt] 
\centering 
\includegraphics[width=150mm,clip]
{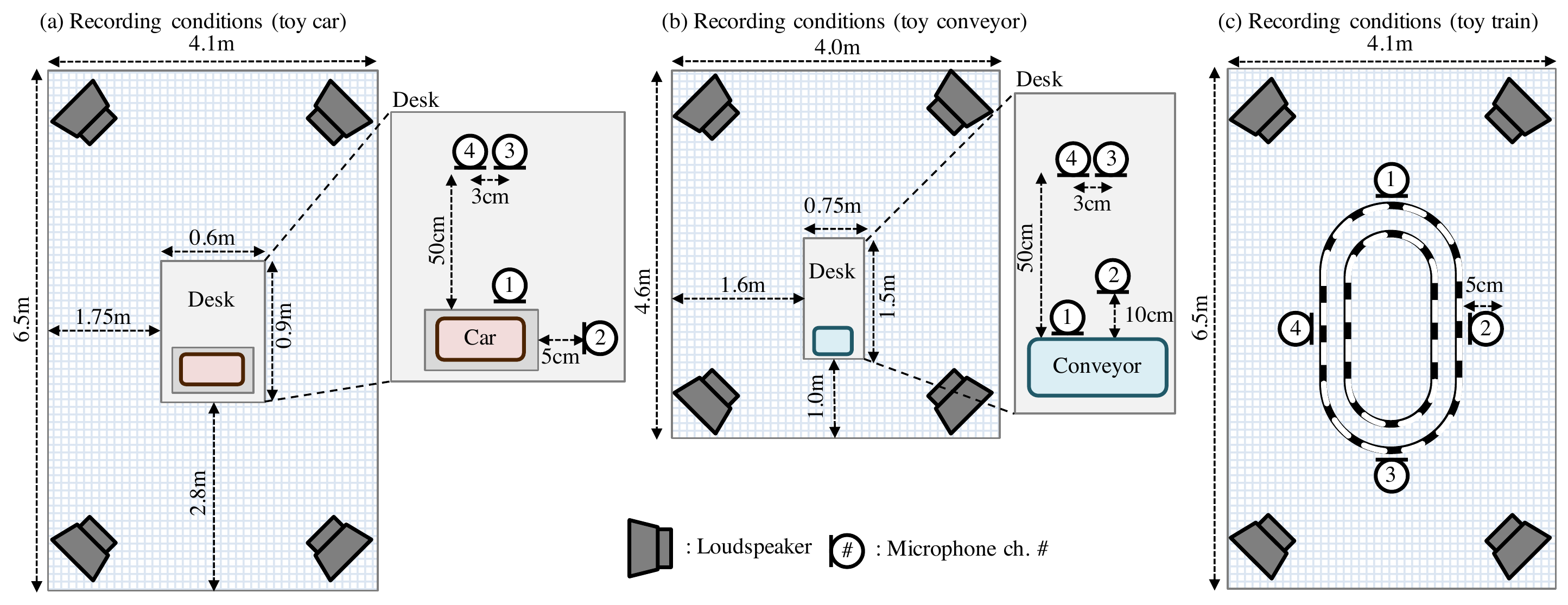} 
\caption{Setup-positions of all equipment (top view)} 
\label{fig:room_overview}
\end{figure*}

The ToyADMOS dataset consists of three sub-datasets for three types of ADMOS tasks. A different toy is used for each task. The name and overview of each sub-dataset are as follows:
\begin{description}
\item[Toy car:] Designed for product-inspection task. A toy car runs on an inspection device. Sound data are collected with four microphones arranged close to the inspection device, as shown in Figs. \ref{fig:mini4wd} (a) and (b), and Fig. \ref{fig:room_overview} (a).
\item[Toy conveyor:] Designed for fault diagnosis of a fixed machine. A toy conveyor is fixed on a desk, and sound data are collected with four microphones. One is fixed on the body of the conveyor, and the other three are placed on the desk, as shown in Fig. \ref{fig:mini4wd} (c) and Fig. \ref{fig:room_overview} (b).
\item[Toy train:] Designed for fault diagnosis of a moving machine. A toy train runs on a railway track. Sound data are collected with four microphones surrounding the track, as shown in Fig. \ref{fig:mini4wd} (d) and Fig. \ref{fig:room_overview} (c).
\end{description}
To collect various normal/anomalous sounds depending on individual differences, operating sounds are recorded using three or four models for each type of toy; these models belong to the same class of toys but have different detailed structures. In the ToyADMOS dataset, we use ``{\bf case}'' as the identifier of each machine. The details  are given in Sec \ref{sec:detail_sub_dataset}.

\begin{figure}[ttt] 
\centering 
\includegraphics[width=85mm,clip]
{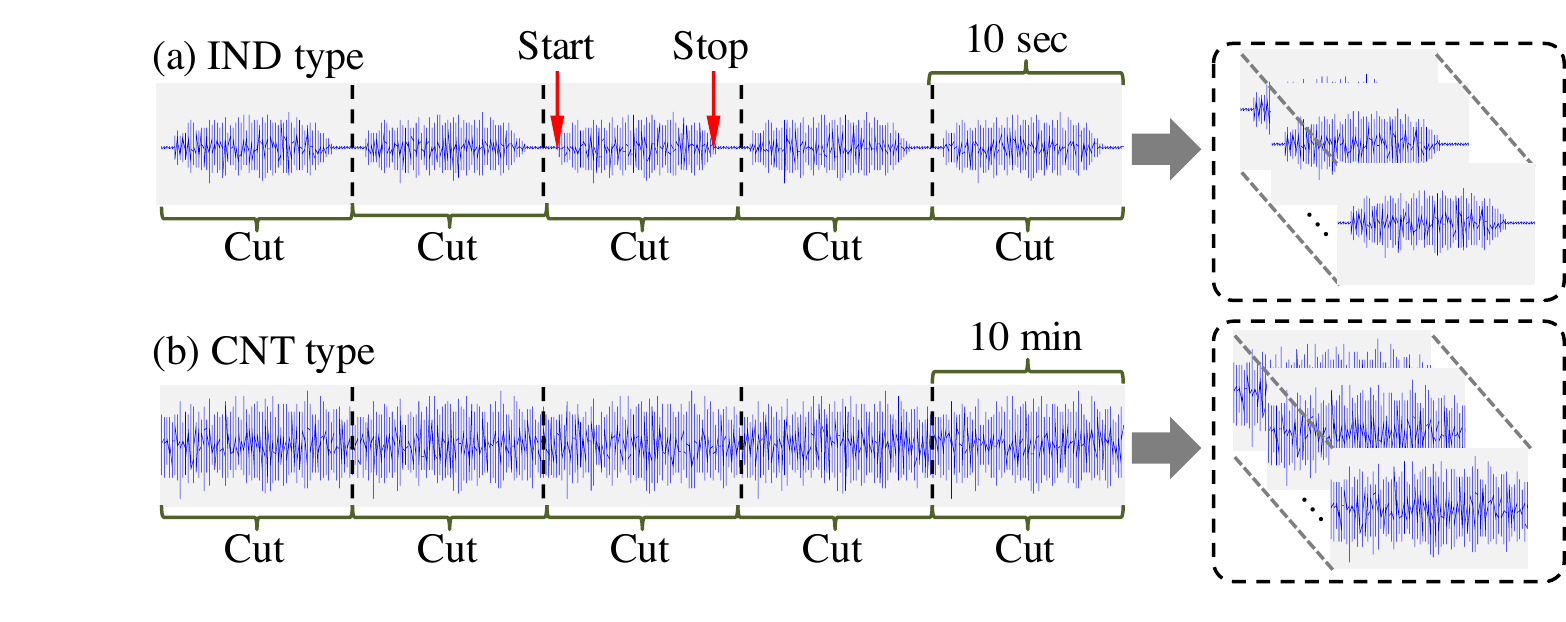} 
\caption{Differences in recording and data processing between individual ({\bf IND}) and continuous ({\bf CNT}) wav-files} 
\label{fig:IndCnt}
\end{figure}

Each sub-dataset consists of three types of sound data: normal, anomalous, and environmental. Their definitions are as follows:
\begin{description}
\item[Normal sound:] Operating sound when the target machine operates normally in accordance with its specifications.
\item[Anomalous sound:] Operating sound when the target machine is made to operate anomalously by deliberately damaging its components or adding extraneous objects.
\item[Environmental noise:] Environmental noise for simulating a factory environment. Noise samples were collected at several locations in an actual factory, such as collision, drilling, pumping, and airbrushing.
These sounds were emitted from four loudspeakers at the corners of each recording room. 
\end{description}
Four omnidirectional microphones (SHURE SM11-CN) were used for collecting these sounds. All sounds were stored as multiple wav-files categorized into two types: individual ({\bf IND}) and continuous ({\bf CNT}). The differences between IND and CNT are shown in Fig. \ref{fig:IndCnt}. IND wav-files contain the operating sounds of the entire operation ({\it i.e.} consisting a starting of a toy to stopping it) in a single wav-file, and each wav-file is approximately 10 sec long. CNT wav-files contain only a some of the operating sounds in a single wave-file; operating sound is recorded continuously and is cut every 10 min. A normal sound consists of both IND- and CNT-files, anomalous sound consists of IND-files, and environmental noise consists of CNT-files. IND and CNT are assumed to be mainly used for training and evaluation, respectively. This is because of the difficulty in collecting IND data in real environments. CNT data can be collected just by recording the operating sound of a working machine, whereas IND-type data collection requires the machine to be started/stopped many times. Thus, IND data collection has a much higher cost than CND data collection, resulting in real-world systems often being trained with CNT datasets.

The main advantage of the ToyADMOS dataset over other datasets \cite{AudioSet,FreeSound} is that it was built under controlled conditions. 
An unsupervised approach is often adopted for ADMOS systems \cite{Marchi_2015,Kawaguchi_2017_MLSP,Koizumi_2019_BU,Kawachi_2018,Kawachi_2019,adaflow} because it is difficult to build an extensive set of anomalous sounds in the real world. 
Therefore, a DNN is trained by using only given normal sound and anomalous sound is defined as ``{\it unknown}'' sounds, 
in contrast to supervised DCASE challenge tasks \cite{DCASE2017,DCASE2016} for detecting ``{\it defined}'' anomalous sounds such as gunshots \cite{Valenzise2007}. 
This definition results in misdetection caused by both a rare normal sound and the difference between the recording condition in training/test dataset. 
Thus, to analyze system performance and/or the cause of misdetection, all normal sounds in dataset need to be collected under the same condition, like as the ToyADMOS dataset.

The limitation of the ToyADMOS dataset is that toy sounds and real machine sounds do not necessarily match exactly. One of the determining factors of machine sounds is the size of the machine. Therefore, the details of the spectral shape of a toy and a real machine sound often differ, even though the time-frequency structure is similar. Thus, we need to reconsider the pre-processing parameters evaluated with the ToyADMOS dataset, such as filterbank parameters, before using it with a real-world ADMOS system.

\section{Details of sub-datasets}
\label{sec:detail_sub_dataset}

\begin{table*}[ttt]
\centering 
\caption{Sound data contained in ToyADMOS dataset}
\begin{tabular}{l|c|c|c} \hline
		 			& Toy car	& Toy conveyor	& Toy train \\ \hline\hline
\# of IND normal sounds per case and channel 		& 1,350 samples		& 1,800 samples		& 1,350 samples	\\
Total hours										& 66	hours			& 60 hours			& 66	hours \\ \hline
\# of CNT normal sounds per case and channel 		& $\approx$ 150 samples 	& at least 124 samples	& 74 samples \\
Total hours										& 135 hours			& 120 hours			& 197 hours \\ \hline
\# of IND anomalous sounds per case and channel 		& $\approx$ 250 samples & 355 samples 		& 270 samples \\ \hline
\# of CNT environmental noise samples per case and channel 	& 72 samples 			& 72 samples 			& 72 samples \\
Total hours										& 48 hours			& 144 hours			& 96 hours \\
\hline
\end{tabular}
\centering 
\caption{List of deliberately damaged parts and their conditions}
\begin{tabular}{l|l|l|l|l|l} \hline
\multicolumn{2}{c}{Toy car} & \multicolumn{2}{|c|}{Toy conveyor} & \multicolumn{2}{c}{Toy train}\\ \hline
\multicolumn{1}{c|}{Parts} & \multicolumn{1}{c|}{Condition} & \multicolumn{1}{c|}{Parts} & \multicolumn{1}{c|}{Condition} & \multicolumn{1}{c|}{Parts} & \multicolumn{1}{c}{Condition} \\ \hline \hline
Shaft 	& - Bent 		& Tension pulley 	& - Excessive tension 	& First Carriage 	& - Chipped wheel axle \\ \hline
Gears 	& - Deformed 	& Tail pulley 		& - Excessive tension 	& Last Carriage 	& - Chipped wheel axle \\
		& - Melted			& 					& - Removed 			& 					& \\ \hline
Tires 	& - Coiled (plastic ribbon) 	& Belt	& - Attached metallic object 1	& Straight railway track	& - Broken \\
		& - Coiled (steel ribbon) 	& 		& - Attached metallic object 2	& 					& - Obstructing stone \\
		& 							& 		& - Attached metallic object 3 & 					& - Disjointed\\ \hline
Voltage & - Over voltage 			& Voltage 	& - Over voltage 		& Curved railway track	& - Broken \\
		 & - Under voltage 			& 			& - Under voltage 		& 					& - Obstructing stone\\
		& 							& 			& 						& 					& - Disjointed \\
\hline
\end{tabular}
\label{table:anm_list}
\end{table*}

\subsection{Toy-car sub-dataset}
We assumed a product inspection task and taken up the task of detecting anomalous sounds from the running sound of a toy car on an inspection device, as shown in Fig. \ref{fig:mini4wd} (b). A toy car called ``mini 4WD'', the four tires of which are driven by a small motor through gears and a shaft, was used as a miniature car machine, as shown in Fig. \ref{fig:mini4wd} (a). The motor and a stabilized power supply were connected, and running sounds on an inspection device were recorded with four microphones. The inspection device, microphones, and loudspeakers in the recording room were arranged as shown in Fig. \ref{fig:room_overview} (a). 

Each ``case'' of the toy car was designed as the combination of two types of motors and bearings; thus, the number of cases was four. Each wav-file of IND normal and anomalous sounds was 11 sec long, and 1,350 IND samples were recorded in each case and channel. The total number of hours of IND normal sounds is 66. Approximately 150 CNT samples were recorded in each case and channel. Note that to reduce the motor load, a 10-min break was given per 10-min operation of the motor. Thus, the total number of hours of CNT normal sounds is 135, which is half the total length of CNT-files. Anomalous sounds were generated by deliberately damaging the shaft, gears, tires, and voltage, as shown in Table \ref{table:anm_list}. In total, 250 samples of anomalous sounds were recorded with these combinations (53 patterns) in each case and channel. Since the microphones were positioned the same in all cases, 12 hours of environmental noise were recorded only once.

\subsection{Toy-conveyor sub-dataset}
We assumed a fault diagnosis task of a fixed machine task in which anomalous sound are detected from the operating sound of a toy conveyor fixed on a desk. We used a conveyor that transports a small tin toy car by driving a belt using a small built-in small, as shown in Fig. \ref{fig:mini4wd} (c, upper). The channel 1 microphone was placed on the body of the conveyor, and the other microphones were placed on the desk, as shown in Fig. \ref{fig:mini4wd} (c, lower). The desk, microphones, and loudspeakers in the recording room were arranged as shown in Fig. \ref{fig:room_overview} (b).

Three types of conveyors, which were produced by the same manufacturers but had different sizes, were used as ``cases'' of toy conveyors; thus, the number of cases was three. Each wav-file of IND normal and anomalous sounds was 10 sec long, and 1,800 IND samples were recorded in each case and channel. Thus, the total number of hours of IND normal sounds is 60. At least 124 CNT samples were recorded in each case, and the total number of hours of CNT normal sounds is 120. Anomalous sounds were generated by deliberately damaging the tension pulley, trail pulley, and belt and excessively lowering/raising the voltage, as shown in Table \ref{table:anm_list}. In total, 355 samples of anomalous sounds were recorded with these combinations (60 patterns). Since the first microphone was placed on the conveyor, the microphones were positioned differently in each case. Thus, 12 hours of environmental noise were recorded in all cases.

\subsection{Toy-train sub-dataset}
We assumed the use of fault diagnosis of a moving machine task, which detects anomalous sounds from the running sound of a toy train. That is, to detect anomalous sounds, we need to combine the observations of four channels. We used HO-scale (large) and N-scale (small) model railways, which are precisely detailed miniature models of railways, as shown in Fig. \ref{fig:mini4wd} (d). Sound data were collected with four microphones surrounding the railway track. The microphones and loudspeakers in the recording room were arranged as shown in Fig. \ref{fig:room_overview} (c). Note that since the sizes of the HO- and N-scale railways differed, the positions of microphones also differed. The microphone arrangements shown in Figs. \ref{fig:mini4wd} (d) and \ref{fig:room_overview} (c) are for the HO-scale railway. We removed the HO-scale railway and moved the microphones close to the N-scale railway when recording N-scale machine sounds. 

Each ``case'' of a toy train is designed as a combination of two types of trains (commuter and a bullet) and scales (HO-scale and N-scale); thus, the number of cases was four. Each wav-file of IND normal and anomalous sounds was 11 sec long, and 1,350 IND samples were recorded in each case and channel. The total number of hours of IND normal sounds is 66. Seventy-four CNT samples were recorded in each case and channel; thus, the total number of hours of CNT normal sounds is 197. Anomalous sounds were generated by deliberately damaging the first/last carriage and straight/curved railway track, as shown in Table \ref{table:anm_list}. In total, 270 samples of anomalous sounds were recorded with these combinations (54 patterns). Since the microphones were positioned differently in the HO- and N-scale cases, 12 hours of environmental noise were recorded for each case.


\section{Evaluation and benchmark}
\label{sec:usage}
To give a usage and a  sense of the usefulness of the ToyADMOS dataset, we tested a simple baseline system on three sub-datasets. The set of Python codes for training, test, and generating the training/test data are available for download at the same address of the ToyADMOS dataset.

We tested a simple unsupervised-ADS task using each sub-dataset of case 1 on channel 1. Note that, to simplify the experiment, we mixed the observations of channels 1--4 in the toy-train sub-dataset and used as a single channel observation. Randomly selected 1,000 samples of IND normal sound samples were used for training, and the other IND normal and IND anomalous sound samples were used for evaluation. We built both training/test datasets by mixing randomly cropped environmental noise. To control the signal-to-noise ratio, we multiplied 3.16 (+10 dB) by the waveforms of target sounds in the toy-car and toy-conveyor sub-datasets and by the waveforms of noise sounds  in the toy-train sub-dataset. All sounds were downsampled to a sampling rate of 16 kHz.

We used a simple AE as a normal model, and its network architecture was almost the same used in \cite{Koizumi_2019_BU}. 
Each encoder/decoder of the AE has one input fully connected neural network (FCN) layer, four hidden FCN layers, and one output FCN layer. Each hidden layer has 512 hidden units, and the dimensions of the encoder output are 128. The rectified linear unit is used after each FCN layer except the output layer of the decoder. The input vector was a 64-dimensional log-mel-amplitude spectrum, and their before/after 10 time-frames were concatenated to account for previous and future frames. The reconstruction error of the AE was used as the anomaly score, and the parameters of the AE were trained to minimize the anomaly score of normal training samples. We fix the learning rate for the initial 100 epochs and decreased it linearly between 100--200 epochs down to $1/100$, where we start with a learning rate of $10^{-4}$. 
We always concluded the training after 200 epochs.

We calculated the anomaly scores on each time frame of all test wav-files. If the anomaly score exceeded the threshold even for one frame, the wav-file was determined to be anomalous. 
This system gave the area under the receiver operating characteristic curves of 0.874, 0.981, and 0.843 for the toy-car, toy-conveyor, and toy-train sub-datasets, respectively. By analyzing false-negative detections 
({\it i.e.} overlooking), we found the system frequently overlooked ``over-voltage'' sounds in the toy-car sub-dataset and anomalous sounds of ``curved railway track'' in the toy-train sub-dataset. The details of the spectral shape of the normal sounds and over-voltage sounds were slightly different; however, there was almost no change in the amplitude of these sounds. Moreover, the damaged point of the curved railway was far from all four microphones; therefore, the amplitude of the overlooked anomalous sounds was small. These results indicate one research direction in ADMOS using the ToyADMOS dataset: detecting anomalous sounds whose time-frequency structure does not change much from the normal sounds.

\section{Conclusions}
\label{sec:cncl}
We introduced a new dataset called ``ToyADMOS'' designed for use in anomaly detection in machine operating sounds (ADMOS). To build a large-scale dataset for ADMOS use, we collected anomalous operating sounds of miniature machines (toys) by deliberately damaging them. The ToyADMOS dataset and some tutorial Python codes are freely available on the Web, and we hope this dataset can contribute to advancing research into anomaly detection in sounds.

\vspace{2pt}
{\bf Acknowledgements:} The authors thank Minato Kanazawa at NTT TechnoCross Corporation for his technical assistance in the data collection process.

\clearpage
\bibliographystyle{IEEEbib}
\bibliography{refs}

\end{document}